\documentclass[aps,prl,groupedaddress,showpacs,twocolumn]{revtex4-1}

\usepackage{braket}
\usepackage{graphicx}
\usepackage{epstopdf}
\usepackage[latin1]{inputenc}
\usepackage{amsmath,amsbsy,amsfonts,amssymb}
\usepackage{epsfig}
\usepackage{hyperref}
\newcommand{\wn}{\mbox{cm$^{-1}$}}

\begin{document}

\title{High accuracy photoassociation of $^{40}$Ca near the ${^3P_1}$+${^1S_0}$ asymptote and its\\
 Zeeman effect}


\author{Max Kahmann$^1$, Eberhard Tiemann$^2$, Oliver Appel$^1$, Uwe Sterr$^1$, and Fritz Riehle$^1$}

\affiliation{$^1$Physikalisch-Technische Bundesanstalt (PTB), Bundesallee 100, 38116 Braunschweig, Germany}

\affiliation{$^2$Institut f\"ur Quantenoptik, Leibniz Universit\"at Hannover, Welfengarten 1, 30167 Hannover, Germany}

\date{\today}

\begin{abstract}
We report on the first measurement of narrow photoassociation lines of $^{40}$Ca near the ${^3P_1}$+${^1S_0}$ asymptote related to the molecular states $^3\Pi_u$ and $^3\Sigma^+ _u$.
The highly accurate binding energies and Zeeman splittings are well described by a coupled channel theoretical model, confirming theoretical predictions of long-range coefficients. 
Our analysis shows that only the inclusion of both energies and Zeeman splittings provides an accurate description of the long-range interaction potentials.      
\end{abstract}


\maketitle


Recently, alkaline earth elements like Mg, Ca, Sr and isoelectronic systems like Yb have attracted broad attention due to their narrow intercombination transitions with various possible applications in physics and technology. These include utilization in optical atomic clocks \cite{fal11, lem09}, quantum computation \cite{yi08}, precision measurements \cite{kot09}, determination of molecular potentials with high accuracy \cite{jon06}, investigation of scattering processes \cite{vog07}, or the production of ultra cold molecules in the ground state \cite{rei12} with the prospect of studying ultracold chemistry. The precise determination of the long range interaction between atoms in the corresponding states by photoassociation spectroscopy is a prerequisite for many applications. 

The first narrow line photoassociation investigations of alkaline earth (like) elements have been performed at the $^1$S$_0 - ^3$P$_1$ transition in Yb \cite{kuw99} (atomic line width $\Gamma/2\pi=182$~kHz) determining scattering lengths and potentials \cite{kit08} and studying molecule formation \cite{kat12}. For the alkaline earth element Sr ($\Gamma/2\pi$=7~kHz) narrow line photoassociation has been observed \cite{zel06,mar08e,ste12}. 

In our investigation with $^{40}$Ca ($\Gamma/2\pi=374(9)$~Hz \cite{deg05a}) we enter a regime where the long range potentials are dominated by the van der Waals ($C_6$) interaction. As has been pointed out by \cite{ciu04} due to the similar van der Waals coefficients in the involved ground and excited state potentials of Ca$_2$ the photoassociation (PA) differs considerably from other elements with significant dipole interaction like Sr and Yb and it leads to a large probability for spontaneous decay to bound ground state molecular levels \cite{ciu04} and does not allow to apply the reflexion approximation \cite{boi00} for the calculation of line intensities.  

\begin{figure}
\includegraphics[width=0.9\columnwidth]{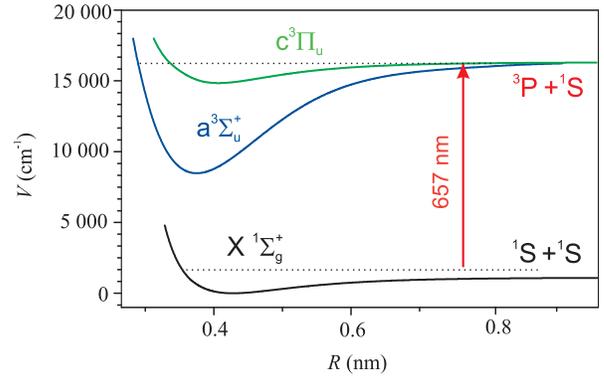}
\caption{Relevant Hund's case (a) molecular potential energy curves of Ca$_2$ as a function of internuclear separation $R$ neglecting spin-orbit interaction. Photoassociation is performed near the ${^1S_0} + {^3P_1}$ asymptote ($\lambda=657$~nm).}
\label{Fig:potentiale}
\end{figure}

The small $C_3$ coefficient (dipole-dipole interaction) in Ca and the dominant $C_6$ interaction result in two excited long range attractive potentials $^3\Pi_{u}$ and $^3\Sigma^+ _{u}$ (Fig. \ref{Fig:potentiale}). This potential scheme differs from all other so far investigated PAs of homonuclear molecules. The molecular states within these similar potentials can effectively couple by rotational and spin-orbit interaction which leads to a striking dependence of the molecular g factor on the rovibrational levels. 

In our experiment bosonic $^{40}$Ca atoms are trapped and cooled in two subsequent stages of magneto-optical trapping \cite{deg05a} and loaded into a crossed optical dipole trap with one horizontal and one tilted beam \cite{kra09} at a temperature of $12~\mu$K. After further evaporation by lowering the trap depth about $N=150~000$ atoms are prepared at a temperature of $T\approx 1~\mu$K and a peak density of $\rho\approx 1.1\cdot10^{19}~$m$^{-3}$. 

The $657$~nm PA laser with a line width below $2$~kHz can be precisely tuned relative to an ultra stable reference resonator \cite{naz06}. For each PA line the eigenfrequency of the resonator was determined with respect to the atomic ${^1S_0}-{^3P_1}$ transition frequency $\nu_{\rm atom}$ \cite{supp}. 

The trapped atoms are irradiated by the PA laser for typically $1$~s and with an intensity of up to $150$~W/cm$^2$.
During PA excitation a homogeneous magnetic field of $B=0.285(7)$~mT, calibrated by the atomic Zeeman splitting \cite{bev98}, is applied.	
The trap loss caused by photoassociation as a function of the laser frequency is observed from absorption images of the atomic cloud. 

\begin{figure}
 
\includegraphics[width=1\columnwidth]{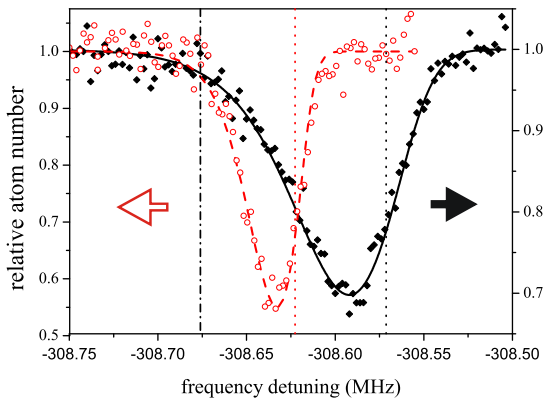}
\caption{Trap loss as function of frequency detuning $\nu - \nu_{\rm atom}$ for the weakest bound $(v'=-1)$ state of the $c^3\Pi_u~(0^+)$ potential. The dipole trap is operated at powers of $0.6$~W in the horizontal beam and $1.6$~W in the tilted beam (full circles) and $0.3$~W and $0.8$~W (open circles), respectively, at a PA intensity of $6$~mW/cm$^2$. The black full line and the red dashed line show fits (see text) for a temperature of $T=1.0~\mu$K and $T=0.5~\mu$K, respectively. The dotted lines indicate the calculated resonance frequencies at $T=0$ and the dash-dotted line the resonance frequency extrapolated to zero trap depth.}
\label{Fig:asymlinie}
\end{figure}

 At the low temperature of our experiment only s-wave scattering is expected for bosons in the ground state and thus  only molecular states with total angular momentum $J=1$ and negative parity can be excited. Fig. \ref{Fig:asymlinie} shows the $M_J=0$ PA resonance for the weakest bound state $v'=-1$ of the $c^3\Pi_u~(0^+)$ potential for two different beam powers of the dipole trap.

We describe the trap loss data as a pure two body loss $\dot{n}=-K n^2$ for density $n$ and with a loss coefficient $K$ given as a sum of Gaussians weighted and shifted according the thermal distribution \cite{jon99}. The two fitted lines in Fig. \ref{Fig:asymlinie} represent a temperature of $1.0~\mu$K and $0.5~\mu$K and a Gaussian Full With at Half Maximum (FWHM) of $42$~kHz and $18$~kHz, respectively. These temperatures agree with the temperatures estimated from the trap depths. Additionally performed time-of-flight measurements for the case of the deep trap yielded $1.1(1)~\mu$K. These temperatures correspond to a Doppler broadened Gaussian with FWHM of $36$~kHz and $25$~kHz, respectively, in fair agreement with the fit. This Doppler width is more than an order of magnitude larger than the molecular natural line width ($ \approx 2\Gamma({^3P_1})/2\pi=748$~Hz) which justifies the choice of Gaussians rather than Lorentzians in modeling the line shape. In difference to other PA measurements \cite{zel06,the04} no phenomenological linewidth broadening factor is introduced.

From the fit we obtain the extrapolation to the position of the resonances at $T=0$, which is indicated in Fig. \ref{Fig:asymlinie} by the dotted lines. Besides the thermal shift the measured energies of the PA resonances are also shifted by the ac-Stark shift from the dipole trap laser beams, operating at a wavelength of $1030$~nm. The unperturbed line position (dash-dotted line) was determined by linear extrapolation to zero dipole-trap depth.

To derive the exact binding energies $h \Delta_{\rm b}$ with respect to the ${^1S_0} - {^3P_1}$ asymptote from such curves we include additional corrections due to molecular photon recoil shift and ac-Stark shift from the PA laser \cite{supp}. The shift from the PA laser was linearly extrapolated to zero intensity from measurements at different PA intensities. The PA induced lightshift for the case in Fig. \ref{Fig:asymlinie} for $6$~mW/cm$^2$ was estimated to be less than $1$~kHz.

We have measured transitions between the ground state continuum of atom pairs to the three weakest bound vibrational states of the two excited \textit{ungerade} potentials (Fig. \ref{Fig:potentiale}), denoted $c^3\Pi_u$ and $a^3\Sigma^+ _u$ in Hund's coupling case (a), which correlate to the asymptote $^1S$+${^3P_1}$ as $c0^+_u$ and the strongly mixed $(a,c) 1_u$, respectively. The binding energy was determined by fitting the loss coefficient $K$ by a single Gaussian instead of the detailed evaluation for Fig. \ref{Fig:asymlinie}. We further assume a linear dependency of the thermal energy shift on the trap depth. Thus the thermal shift as well as the light shift were corrected by the linear extrapolation to zero dipole trap laser intensities.

\begin{figure}
 
\includegraphics[width=1\columnwidth]{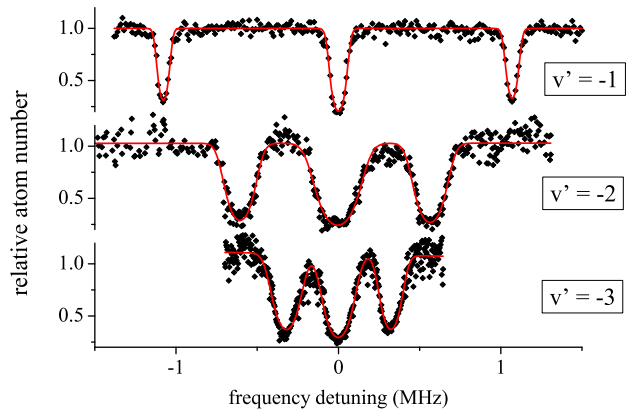}
\caption{Trap loss as a function of frequency relative to the $M_J=0$ component showing the Zeeman triplet for the three weakest bound molecular states in the $c^3\Pi_{u}$ ($0^+$) potential for a magnetic field of $B=0.285(7)$~mT. The intensities of the PA laser were $I(v'=-1)=3.7$~W/cm$^2$,  $I(v'=-2)=57$~W/cm$^2$, $I(v'=-3)=127$~W/cm$^2$. The red solid curves are simulated profiles, see text.}
\label{Fig:zeeman}
\end{figure}

For each of these lines we applied individual corrections as described above. The corrected binding energies $h \Delta_{\rm b}$ are given in table \ref{measurements} together with derived uncertainties as detailed in \cite{supp}. 

From the observed Zeeman splitting (Fig. \ref{Fig:zeeman}) the molecular g factor was determined (see table \ref{measurements}). The magnetic field is so small, that only a linear Zeeman effect is expected. Observed small and unsystematic asymmetries of the Zeeman splittings in particular triplets were treated as contribution to the uncertainties. In our setup the PA laser beam is perpendicular to the magnetic field, thus we can not distinguish between $\Delta M = \pm1$ transitions and only the absolute values of the g factors could be experimentally determined.  The fact that we observe significant Zeeman-splitting in all states also for $\Omega =0$ requires a theoretical model which takes into account the coupling by spin-orbit and rotational interaction.

\begin{table}[t]
\centering
\begin{tabular}{|c|c|c|c|c|c|c|}
\hline
\multicolumn{2}{|c|}{~level~} & $\Delta_{\rm b}^{\rm calc}$ & $\Delta_{\rm b}^{\rm exp}$ & $\delta$ & $g^{\rm calc}$ & $|g^{\rm exp}|$  \\
$~v'$ & $\Omega$ & GHz & GHz & kHz & &  \\
\hline
-1 & 0 & -0.308672 & ~-0.308670~(7) & 2 & -0.273 & 0.276~(1)  \\
\hline
-1 & 1 & -0.982995 & ~-0.982994~(6) &  1 & ~1.069 & 1.074~(4)  \\
\hline
-2 & 0 & -4.649199 & ~-4.649203~(22) & -4 & -0.147 & 0.147~(6)  \\
\hline
-2 & 1 & -7.411944 & ~-7.411946~(12) & -2 & ~0.902 & 0.901~(27)  \\
\hline
-3 & 0 & -17.857274 & -17.857274~(7) & 0 & -0.079 & 0.081~(3)  \\
\hline
-3 & 1 & -24.539446 & -24.539446~(7) & 0 & ~0.831 & 0.843~(26)  \\
\hline
-4 & 0 & -44.735  &  & &  -0.04 &   \\
\hline
-4 & 1  & -57.860  &  & & ~0.80 &   \\
\hline
\end{tabular}
\caption{Measured and calculated binding energies $h \Delta_{\rm b}$ (with $J=1$), their difference $\delta$ and molecular g factors of the $c0_u^+$ and $(a,c)1 _u$ states with vibrational state $v'$ counted from the asymptote.}
\label{measurements}
\end{table}

Our model is similar to the one of Ciurylo \textit{et al.} \cite{ciu04}. Their model spans the Hilbert space of molecular states resulting from atomic pairs in quantum states ${^1S_0} + {^3P_j}$ and ${^1S_0}+{^1P_1}$ with $j=1$ and $2$. We reduce it to ${^1S_0}+{^3P_j}$, because the singlet-triplet coupling in Ca is much weaker than in Sr, but we include this coupling by setting up weak effective dipole-dipole terms in the long range potentials, the magnitudes of which are estimated from the radiation life time of ${^3P_1}$ and the transition frequency of the intercombination line. Because s-wave photoassociation from ${^1S_0}+{^1S_0}$ will only excite molecular levels with total angular momentum $J=1$ and parity $-$, we consider the following three molecular states given in the notation of Hund's coupling case (e) 

$\ket{(L,S)j,l,J}$: 

$\ket{(1,1)1,0,1}$, $\ket{(1,1)1,2,1}$ and $\ket{(1,1)2,2,1}$.
 
\noindent Here, $l$ describes the overall rotation of the atom pair and the other symbols have their conventional meaning. The related molecular potentials in Hund's case (a) have the labels $a^3\Sigma^+_u$ and $c^3\Pi_u$, where only the $u$ symmetry is relevant for the present photoassociation. The molecular potentials (see Fig. \ref{Fig:potentiale}) are constructed from known spectroscopic results \cite{all05} and extensions for long range with $R>R_o = 1.2$~nm:
\begin{eqnarray}
\label{eq:potential0}
V_{a}=& -\frac{-4C_{3}^{(0)}}{R^3} -\frac{C_{6}^a}{R^6}-\frac{C_{8}^a}{R^8}  \\
V_{c}=& -\frac{2C_{3}^{(0)}}{R^3}-\frac{C_{6}^c}{R^6}-\frac{C_{8}^c}{R^8}.
\label{eq:potential1}
\end{eqnarray}

The resonant dipole-dipole term $C_3^{(0)}$ is estimated according to \cite{ciu04}\footnote{There is an opposite sign definition in that paper for the dipole-dipole term compared to the conventional usage, which is applied through the present paper} from the lifetime of state ${^3P_1}$ $\tau=0.426(10)$~ms \cite{deg05a} and the transition frequency $15210.064$  \wn~ \cite{deg05a} of ${^1S_0}\rightarrow  {^3P_1}$ giving  $C_3^{(0)}=10.72 $ \wn\AA$^3$. The full Hamiltonian includes the spin-orbit interaction and the molecular potentials, which are transformed from case (a) to case (e) using the transformation matrix as given in the appendix of \cite{ciu04}. The three relevant basis vectors in Hund's case (a)  $\ket{\Lambda,S,\Sigma,J}$ are:

 c$\ket{1,1,-1,1}$, c$\ket{1,1,0,1}$ and a$\ket{0,1,1,1}$.

\noindent The first two states will be mixed by Coriolis coupling and the last two by spin-orbit interaction. Because of the general dipole-dipole term for state c, the Hamiltonian will contain such contribution also for the basis vector c$\ket{1,1,0,1}$, which correlates to the atomic pair ${^1S_0}+{^3P_2}$ where this term should actually vanish. We neglect this contribution because of the large spin-orbit interaction compared to this artificial dipole-dipole term. The resulting amount of energy in the eigenstates correlating to ${^1S_0}+{^3P_1}$ is only in the order of the ex\-pe\-ri\-men\-tal uncertainty of the energy measurements.
The eigenenergies and eigenvectors are determined by diagonalizing the Hamiltonian in discrete variable representation (DVR) with appropriate mapping functions for reducing the number of grid points on the internuclear axis.  

The Zeeman energy can be well approximated in Hund's case (e), neglecting the rotational Zeeman energy, which is very small for a homonuclear molecule with completely balanced charge distribution \cite{tow55}. The matrix element of the Zeeman operator for a magnetic field $B_0$ in space fixed direction $z$ can be written:

\begin{equation}
\begin{split}
\bra{j,l,J,M}\mu_Bg_jB_0j_z\ket{j,l,J,M}=\\
\mu_B\cdot g_j\cdot B_0 \sum_{m_j,m_l}cg(j,l,m_j,m_l;JM)^2\cdot m_j ,
\label{eq:Zeeman}
\end{split}
\end{equation}

\noindent where $\mu_B$ is Bohr's magneton,  $g_j$ the atomic g factor, and the function $cg()$ the Clebsch-Gordan coefficient for the coupling of $j$ with $l$ to the total angular momentum $J$. The Clebsch-Gordan coefficient projects out the appropriate magnetic moment for the rotating pair in state $l$. The total Zeeman energy is calculated by the sum over the needed matrix elements times the probability of the corresponding basis state in Hund's case (e) for the considered eigenvector. We neglect any quadratic Zeeman effect, because the applied field is too low to give a significant contribution.

\begin{table}[t]
\centering
\begin{tabular}{|c||c|c|c||c|c|c|}
\hline
state & \multicolumn{3}{|c||}{  $C_6$  } & \multicolumn{2}{|c|}{ $C_8$}\\
\hline
 & \multicolumn{3}{|c||}{  $10^{7}$ \wn \AA$^6$  } & \multicolumn{2}{|c|}{ $10^{9}$ \wn \AA$^8$}\\
\hline
 & exp. & \cite{ciu04}& \cite{mit08} & exp. & \cite{mit08}   \\
\hline
 $c^3\Pi_u$&1.199&1.187 & 1.226 &0.289&0.266  \\
 $a^3\Sigma_u$& 1.353 &1.313  & 1.358 &0.818&1.057   \\
\hline
\end{tabular}
\caption{Derived long range parameters at the asymptote  ${^1S_0}+{^3P_j}$ and comparison with theoretical results from Mitroy and Zhang  \cite{mit08} and Ciurylo \textit{et al.} \cite{ciu04}.}
\label{Ci}
\end{table}

The energies of the observed six PA resonances and their molecular g factors were fitted simultaneously. A non-linear least squares fit routine varies the long range parameters $C_6$ and $C_8$ and the slope of the short range repulsive branch to get the proper phase of the long range wave function. The $C_3^{(0)}$ coefficient was held fixed at the value given above. In total, six parameters were adjusted for twelve observables. Only the six energies are very precise and thus we expect significant correlations between these parameters. The fit results are given in table \ref{measurements}, where one sees that the energies are reproduced almost exactly, which reflects the case of six highly precise energies fitted by six parameters, the additional data on the g factors look like a byproduct only. But during the fitting procedure we found solutions, which describe the energies as precisely as given in the table while the g factors were significantly off. To force a fit for modeling also the g factors, we increased the weight of those by up to a factor 10000 for some iteration steps and used the proper weighting by the experimental uncertainty only at the end. The derived long range parameters are shown in table \ref{Ci} and compared with earlier theoretical results. The $C_6$ coefficients agree to about 2\% and have values between the theoretical estimates. The $C_8$ values show deviations up to 20\%. We believe that this deviation between experiment and theory is not really a discrepancy and we would attribute it partly to unresolved correlations between the experimentally derived values. But we point out, that this fairly close agreement to theory was only obtained, when fitting properly the g factors.

We studied the internal dependencies on the long range parameters by setting different cases of $C_i$. First we checked the significance of the dipole-dipole term and set the $C_3^{(0)}$ value to zero in a fit. We found a set of parameters which described the energies very well but failed for the observed molecular g factors. Thus despite the small value of  $C_3^{(0)}$, it is necessary for getting the proper coupling between the states in Hund's case (e). From the many trial fits with combinations of $C_6$ and $C_8$ depending on the chosen magnitude of $C_3^{(0)}$ we see the statistical spread of the derived coefficients with their correlation. We conclude that the accuracy of the derived $C_6$ values should be about 5\% and those for the $C_8$ about 20\%, but the total long range function will be much better in the range $R>1.6~$nm.

The Zeeman effect is well described by our model including its strong dependence on the quantum number $\Omega$ and on the magnitude of the binding energy. It also gives the sign of the g factor. During the evaluation of the data we found fits in which within the series of observed levels a level with $\Omega=1$ belonging to the asymptote  ${^1S_0}+{^3P_2}$ appeared. The energies were still well described, but the g factors immediately indicate that this was a wrong solution. Thus having Zeeman data gives a very important insight into the kind of superposition of the basis vectors. As known from many examples in physics, energies are less sensitive to the composition of eigenvectors than magnetic or electric moments. The theoretical g factors are slightly smaller than the experimental ones. Such deviation could indicate that a coupling to an additional electronic state is missing in the theoretical model. This could be a state from asymptote ${^1S_0}+{^1P_1}$ or ${^1S_0}+{^1D_2}$. The latter case is interesting because this coupling was already studied by Allard \textit{et al.} \cite{all05} in the deeply bound region. Here the combination of both experiments would be of scientific value.

Table \ref{measurements} also includes an extrapolation to larger binding energies, but we limited this to the next missing member of each series, because the correlation in the long range parameters prohibits reliable long extensions. The energies given should be good up to few ten MHz and the g factors to $\pm0.02$. 

In conclusion we have presented the first measurement of the narrow photoassociation lines of $^{40}$Ca near the ${^3P_1}$+${^1S_0}$ asymptote and 
we described the observed energy levels and low-field Zeeman splittings by a theo\-re\-ti\-cal model, including rotational and spin-orbit interactions.
The coupled channel calculation indicates that for the exact experimental determination of the long range potential parameters the measurement of the g factors is essential. 

The achieved accuracy is already more than one order of magnitude higher than PA measurements for Yb \cite{toj06} and several times higher compared to PA measurements for Sr \cite{zel06}. 
In contrast to these elements in Ca the current width of the PA resonance is still well above the natural line width. Thus the accuracy can be significantly improved by going to lower temperatures and thereby reducing the Doppler broadening.
A temperature of $150~$nK has already been obtained by evaporative cooling leading to a $^{40}$Ca Bose-Einstein-Condensate \cite{kra09}. 
The determination of binding energies with less than $1$~kHz uncertainty, possible under these conditions, and/or measurements of deeper bound states will further establish Ca as a prototype system to test theoretical predictions for the long range potentials and 
help to develop more complete coupled channel models, where the coupling to other electronic states, e.g. correlating to the ${^1S_0}+{^1P_1}$ asymptote, can be considered.

The knowledge of the PA resonances allows to generate ultracold molecules in the electronic ground state by two photon PA or spontaneous decay. Such decay into bound ground state molecular levels is strongly enhanced due to the similarity of ground state and excited state potentials \cite{ciu04,koc08}.

The narrow line width of alkaline earth elements is expected to solve the loss problem \cite{ciu05} associated with the use of optical Feshbach resonances for modification of the scattering length \cite{fed96,the04}.
Compared to first experiments with Sr \cite{bla11,yan13} and Yb \cite{eno08a}, the even narrower linewidth of Ca lets us expect smaller losses, which enables a wide range of applications, like highly flexible modification of the particle interaction on short time and length scales through the scattering length \cite{rap12,yam10}. 

This work was supported by Deutsche Forschungsgemeinschaft (DFG) through the {\it Center of Quantum Engineering and Space-Time Research (QUEST)} of the Leibniz Universit\"at Hannover and through the Research Training Group 1729 {\it Fundamentals and Applications of Ultra Cold Matter}.
We thank Roman Ciurylo for helpful discussions and the experimental support of Evgenij Pachomov, Stephan Schulz and Sebastian Kraft is gratefully acknowledged.


%

\newpage

This supplement presents the details of modeling of trap loss, the determination of the binding energies $h \Delta_b$ and g factors of the photoassociation (PA) resonances relative to the atomic asymptote and the corresponding measurement uncertainties.  


\section{Trap loss description}

To determine the line center from the observed atom-loss spectra, we model
 the evolution of the atomic density $n$ during photoassociation by
\begin{equation}
\dot{n}=-K(\Delta , I, T)~n^2 ,
\end{equation}
where $K$ is the PA loss rate coefficient which is a function of the frequency detuning $\Delta$ from the PA resonance, the PA spectroscopy laser intensity $I$ and the atomic temperature $T$. 
In this equation the contribution of three-body losses \cite{kra09} (proportional to $n^3$) were neglected, 
as this effect leads to less than 20\% atom loss during the duration of PA of $1$~s.
Losses from collisions with background gas (proportional to $n$) were also neglected, since the corresponding lifetime in the crossed dipole trap was measured to be $\tau \approx 56$~s, which leads to insignificant atom loss during the duration of PA.

Assuming constant trap size during PA, the atomic density is proportional to the total atom number $N$. 
Thus we use the solution of the differential equation 
\begin{equation}
	\dot{N} = -\beta(\Delta, I, T)~N^2, 
\end{equation}
where the coefficient $\beta \propto K$ depends on the trap geometry and the temperature $T$. The solution of this differential equation is given by
\begin{equation}
N(t)=\frac{N_0}{1+t\;\beta(\Delta) N_0},
\label{eq:de}
\end{equation}
where $t$ is the PA exposure time and $N_0$ the initial atom number, 
which we take as the atom number far of resonance ($\beta=0$).

\section{PA frequency corrections}

To determine the binding energy $h \Delta_{\rm b}$ from the frequency of the unperturbed PA resonance relative to the atomic asymptote several corrections were taken into account: 
the photon recoil shift for the Ca$_2$ molecule and the Ca atom, 
the thermal shift due to the initial kinetic energy from the relative motion of the colliding atoms of about $3/2\cdot k_{\rm B}T/h$ \cite{jon99}, 
and light shifts from the dipole trap and the PA laser.

In the one photon excitation from a single laser beam with wave vector $k$, the absorbed photon frequency is bigger than the atomic or molecular transition frequency due to the additional kinetic energy of the recoiling particle with mass M of
\begin{equation}
 h \Delta_{\rm rec} = \hbar^2 k^2/2M , 
\end{equation}
which amounts to $\Delta_{\rm rec} = 11.55~$kHz for a $^{40}$Ca atom and $5.78~$kHz for a Ca$_2$ molecule. 

For the PA lines, the molecular light shift of the dipole trap and the shift due to thermal energy was corrected simultaneously by varying the dipole trap depth $U$ by a factor of two. 
We assume a light shift proportional to the dipole laser intensity and thus to $U$.
Since in our experimental conditions due to evaporation the temperature is proportional to $U$, also the thermal energy shift shows the same dependence. Time of flight (TOF) measurements and a line fit using a detailed model including the thermal broadening (see figure 2) both confirm a temperature proportional to $U$ to an uncertainty of $0.1~\mu$K.  

Thus describing the observed loss at the two trap depths by single Gaussian-shaped loss coefficients $\beta(\Delta)$, the unperturbed resonance ($T=0$ and no light shift from dipole trap) was obtained by linear extrapolation to $U=0$. 
For the first measured PA line this simplified analysis was compared to the complex fit (see Fig. 2) which leads to the same unperturbed resonance within $4$~kHz and therefore below the fit uncertainty.

The uncertainty of this correction is due to statistical uncertainties in the frequency values from the Gaussian fit to the measurement data used for the extrapolation. 
The non-linearity of the trap laser intensity variation was checked to be less than $1\%$ and was therefore not taken into account separately.
Furthermore the uncertainty in the two temperatures of $0.1~\mu$K leads to an additionally frequency uncertainity of $4.5$~kHz for the extrapolated value at $T=0$.

For the PA resonances $v'=-1$ at $309$~MHz and $983$~MHz (see Table I in the main text) the power was $0.6$~W in the horizontal dipole trap beam and $1.6$~W in the tilted beam and $0.3$~W and $0.8$~W for half trap depth, respectively. For deeper bound PA lines the dipole trap power was $0.6$~W in the horizontal beam and $1.0$~W in the tilted beam and $0.3$~W and $0.5$~W for half trap depth, respectively. The determined molecular light shifts including contribution due to kinetic collision energy and the fit uncertainty are shown in table \ref{atomlight}. 

\begin{table*}[hbt]
\centering
\begin{tabular}{|c|c|c|c|c|c|c|}
\hline
state $v'$ & -1 & -1 & -2  & -2    & -3 & -3 \\
$\Omega$ & 0 & 1 & 0  & 1    & 0 & 1 \\
\hline
\hline
molecular recoil shift& -6  & -6 & -6  & -6   & -6  & -6  \\
\hline
atomic recoil shift& 12  & 12 & 12  & 12   & 12  & 12  \\
\hline
PA light shift& 0 (3)  & 0 (2) & 15 (1)  & 15 (1)   & 17 (2)  & 9 (3)  \\
\hline
molecular light shift by dipole trap and thermal shift & -80 (5)  & -82 (5)  & -26 (6)  & -36 (11)   & -42 (5)  & -60 (5)  \\
\hline
atomic light shift by dipole trap& 122 (2) & 122 (2)  & 110 (21)   & 0 (5)   & 0 (4)  & 0 (5) \\
\hline
\hline
total uncertainty & 7 & 6  & 22  & 12    & 7  & 7  \\
\hline
\end{tabular}
\caption{Corrections with uncertainties in (kHz) applied to the measured line positions at full dipole trap power for the $c0_u^+$ and $(a,c)1_u$ states to obtain the binding energies $h \Delta_{\rm b}$. The quantum number $v'$ determines the vibrational state counted from the asymptote.}
\label{atomlight}
\end{table*}

In addition to the ac-Stark shift due to the dipole trap the PA lines are also influenced by the ac-Stark shift due to the PA spectroscopy laser. In order to determine this ac-Stark shift the PA resonances were measured with different PA laser intensities. 
We assume a linear dependence of the ac-Stark shift on the light intensity and extrapolate to zero PA laser intensity (table \ref{atomlight}). 
The weakest bound states $v'=-1$ were measured at low PA intensities, where no light shift due to the PA inducing laser light was observed within the experimental uncertainty.

To compare the PA line position with the atomic transition frequency, 
within one hour in a second step the atomic $^1S_0-^3P_1$ transition was also measured relative to the eigenfrequency of a ULE reference resonator.
Due to technical reasons the atomic line was measured under different conditions in each measurement run. Thus such measurements were repeated for each individual PA measurement.

For the two weakest bound states $v'=-1$ the atomic reference transition is measured in the dipole trap. 
Here we observed the atomic line at half the dipole trap depth and linearly extrapolated to zero trap depth with an uncertainty given by the fit of $2~$kHz.  
As mentioned above the non-linearity of the intensity variation was less than $1\%$ and therefore was not taken into account separately.

The atomic spectrum for the $v'=-2$ state in potential $c0_u^+$ was measured only for a single trap power of $0.6$~W in the horizontal beam and $1.2$~W in the tilted beam. To correct the atomic light shift due to the dipole trap we use the laser beam radii at trap position of $w_{\rm hor} = 36~\mu$m, $w_{\rm tilt} = 77~\mu$m) to calculate the corresponding intensity. The light shift was then corrected using the dependency on the intensity as determined in the $v'=-1$ measurement run.  
Here the total uncertainty is given by the uncertainty of the light shift determination for the states $v'=-1$ plus uncertainty in the beam radii (contributing $3$~kHz) and an additional uncertainty from an estimated 20\% uncertainty in intensity of the dipole trap beams due to possible changes in the alignment of the dipole laser beams (contributing $20$~kHz).

For deeper bound molecular states the eigenfrequency of the ULE resonator relative to the atomic asymptote was measured in absence of the dipole trap and thus no light shift correction was required. 
The uncertainties for these measurements are therefore given by the fit uncertainty of the spectra. (see Table \ref{atomlight})

As the PA lines and the atomic line were measured one after another within less than one hour with respect to an ultrastable reference cavity, 
also the stability of that cavity had to be considered.
Averaging over months we observe a drift of the cavity eigenfrequency due to aging of $34~$mHz/s, which was automatically compensated.
Additional deviations during a day of less than $\pm2$~kHz were determined by previous absolute frequency measurements with a femtosecond frequency comb. Also during repeated measurements of the atomic transition during this work no additional drift was observed within the experimental resolution of $4$~kHz, thus the uncertainty due to incompletely compensated cavity drift is neglected.

\begin{table*}[hbt]
\centering
\begin{tabular}{|c|c|c|c|c|c|c|}
\hline
state $v'$ & -1 & -1 & -2  & -2    & -3 & -3 \\
$\Omega$ & 0 & 1 & 0  & 1    & 0 & 1 \\
\hline
\hline
B-field calibration uncertainty (\%)  & 0.0 & 0.0  & 3.0  & 3.0   & 3.0  & 3.0 \\
\hline
fit uncertainty m=0 (kHz)& 1.4 & 1.1 &  2.0 &  1.7  & 1.6  & 1.2  \\
\hline
fit uncertainty m=+1 (kHz) & 3.2  & 1.3  & 3.0  & 2.5   & 1.6  & 5.2  \\
\hline
 fit uncertainty m=-1 (kHz) & 1.0 & 1.5  & 2.2   & 3.0   & 1.7  & 3.0 \\
\hline
asymmetry (kHz) & 1.0 & 25.0  & 30.0  & 7.0    & 2.0  & 30.0  \\
\hline
\hline
g factor & 0.276 & 1.074  & 0.147  & 0.901    & 0.081  & 0.843  \\
\hline
total g factor uncertainty  & 0.001 & 0.004  & 0.006  & 0.027    & 0.003  & 0.026  \\
\hline

\end{tabular}
\caption{Estimated uncertainties of the Zeeman splitting of the $c0_u^+$ and $(a,c)1 _u$ states. 
The quantum number $v'$ determines the vibrational state counted from the asymptote.}
\label{Zeeman}
\end{table*}

\section{ ${\bf g}$ factor determination}

In order to determine the g factors of each PA resonance additionally to the $M_J=0$ component the $M_J=\pm$1 components were measured. 
Simplified fits using a loss coefficient with a single Gaussian frequency dependence of each Zeeman component were used to determine the line centers. A drift of the initial atom number $N_0$ during the measurement was taken into account by a linear variation of $N_0$.
The magnetic field was ca\-li\-bra\-ted by measuring the atomic Zeeman splitting at a given current through the Helmholtz coil and using the atomic g factor $g_1=1.5010829(28)$ \cite{bev98} of the $^3P_1$ state. 

Two effects mainly contribute to the uncertainty of the molecular g factors for each PA resonance: The uncertainty of the magnetic field and the asymmetry of the splitting (see Table \ref{Zeeman}). 

For the weakest bound states $v'=-1$ we could measure the atomic Zeeman splitting within minutes after measuring the molecular Zeeman splitting. Several measurements during a day indicate that no resolvable change in the field strength on this time scale occurred. Therefore the uncertainty in the field strength was below 0.1\%. 
For deeper bound molecular states it was not possible to observe the atomic Zeeman splitting during the same measurement run. The field strength was therefore calibrated several times on different days by the atomic Zeeman splitting of $5.99(18)~$MHz for the same coil current. 

The asymmetry of the splitting, i.e. the difference between the $M_J = -1, 0$ and the $M_J = 0, 1$ splitting was up to $30~$kHz, which is in the order of 1\% of the total splitting.
Possible mechanisms for the asymmetry could be a quadratic molecular Zeeman effect, tensor and vector light shifts caused by the dipole trap and $M_j$ dependent light shifts from the PA laser.
As the asymmetry did not show a clear systematic behaviour we have used the mean value of the splitting for the determination of the effective g factors and have included the full asymmetry into the uncertainty. 
 
All uncertainties are shown in Table \ref{Zeeman}.


\begin{thebibliography}{34}%
\makeatletter
\providecommand \@ifxundefined [1]{%
 \@ifx{#1\undefined}
}%
\providecommand \@ifnum [1]{%
 \ifnum #1\expandafter \@firstoftwo
 \else \expandafter \@secondoftwo
 \fi
}%
\providecommand \@ifx [1]{%
 \ifx #1\expandafter \@firstoftwo
 \else \expandafter \@secondoftwo
 \fi
}%
\providecommand \natexlab [1]{#1}%
\providecommand \enquote  [1]{``#1''}%
\providecommand \bibnamefont  [1]{#1}%
\providecommand \bibfnamefont [1]{#1}%
\providecommand \citenamefont [1]{#1}%
\providecommand \href@noop [0]{\@secondoftwo}%
\providecommand \href [0]{\begingroup \@sanitize@url \@href}%
\providecommand \@href[1]{\@@startlink{#1}\@@href}%
\providecommand \@@href[1]{\endgroup#1\@@endlink}%
\providecommand \@sanitize@url [0]{\catcode `\\12\catcode `\$12\catcode
  `\&12\catcode `\#12\catcode `\^12\catcode `\_12\catcode `\%12\relax}%
\providecommand \@@startlink[1]{}%
\providecommand \@@endlink[0]{}%
\providecommand \url  [0]{\begingroup\@sanitize@url \@url }%
\providecommand \@url [1]{\endgroup\@href {#1}{\urlprefix }}%
\providecommand \urlprefix  [0]{URL }%
\providecommand \Eprint [0]{\href }%
\providecommand \doibase [0]{http://dx.doi.org/}%
\providecommand \selectlanguage [0]{\@gobble}%
\providecommand \bibinfo  [0]{\@secondoftwo}%
\providecommand \bibfield  [0]{\@secondoftwo}%
\providecommand \translation [1]{[#1]}%
\providecommand \BibitemOpen [0]{}%
\providecommand \bibitemStop [0]{}%
\providecommand \bibitemNoStop [0]{.\EOS\space}%
\providecommand \EOS [0]{\spacefactor3000\relax}%
\providecommand \BibitemShut  [1]{\csname bibitem#1\endcsname}%
\let\auto@bib@innerbib\@empty













\bibitem [{\citenamefont {Falke}\ \emph {et~al.}(2011)\citenamefont {Falke},
  \citenamefont {Schnatz}, \citenamefont {Vellore~Winfred}, \citenamefont
  {Middelmann}, \citenamefont {Vogt}, \citenamefont {Weyers}, \citenamefont
  {Lipphardt}, \citenamefont {Grosche}, \citenamefont {Riehle}, \citenamefont
  {Sterr},\ and\ \citenamefont {Lisdat}}]{fal11}%
  \BibitemOpen
  \bibfield  {author} {\bibinfo {author} {\bibfnamefont {S.}~\bibnamefont
  {Falke}}, \bibinfo {author} {\bibfnamefont {H.}~\bibnamefont {Schnatz}},
  \bibinfo {author} {\bibfnamefont {J.~S.~R.}\ \bibnamefont {Vellore~Winfred}},
  \bibinfo {author} {\bibfnamefont {T.}~\bibnamefont {Middelmann}}, \bibinfo
  {author} {\bibfnamefont {S.}~\bibnamefont {Vogt}}, \bibinfo {author}
  {\bibfnamefont {S.}~\bibnamefont {Weyers}}, \bibinfo {author} {\bibfnamefont
  {B.}~\bibnamefont {Lipphardt}}, \bibinfo {author} {\bibfnamefont
  {G.}~\bibnamefont {Grosche}}, \bibinfo {author} {\bibfnamefont
  {F.}~\bibnamefont {Riehle}}, \bibinfo {author} {\bibfnamefont
  {U.}~\bibnamefont {Sterr}}, \ and\ \bibinfo {author} {\bibfnamefont
  {C.}~\bibnamefont {Lisdat}},\ }\href {\doibase
  doi:10.1088/0026-1394/48/5/022} {\bibfield  {journal} {\bibinfo  {journal}
  {Metrologia}\ }\textbf {\bibinfo {volume} {48}},\ \bibinfo {pages} {399}
  (\bibinfo {year} {2011})}\BibitemShut {NoStop}%
\bibitem [{\citenamefont {Lemke}\ \emph {et~al.}(2009)\citenamefont {Lemke},
  \citenamefont {Ludlow}, \citenamefont {Barber}, \citenamefont {Fortier},
  \citenamefont {Diddams}, \citenamefont {Jiang}, \citenamefont {Jefferts},
  \citenamefont {Heavner}, \citenamefont {Parker},\ and\ \citenamefont
  {Oates}}]{lem09}%
  \BibitemOpen
  \bibfield  {author} {\bibinfo {author} {\bibfnamefont {N.~D.}\ \bibnamefont
  {Lemke}}, \bibinfo {author} {\bibfnamefont {A.~D.}\ \bibnamefont {Ludlow}},
  \bibinfo {author} {\bibfnamefont {Z.~W.}\ \bibnamefont {Barber}}, \bibinfo
  {author} {\bibfnamefont {T.~M.}\ \bibnamefont {Fortier}}, \bibinfo {author}
  {\bibfnamefont {S.~A.}\ \bibnamefont {Diddams}}, \bibinfo {author}
  {\bibfnamefont {Y.}~\bibnamefont {Jiang}}, \bibinfo {author} {\bibfnamefont
  {S.~R.}\ \bibnamefont {Jefferts}}, \bibinfo {author} {\bibfnamefont {T.~P.}\
  \bibnamefont {Heavner}}, \bibinfo {author} {\bibfnamefont {T.~E.}\
  \bibnamefont {Parker}}, \ and\ \bibinfo {author} {\bibfnamefont {C.~W.}\
  \bibnamefont {Oates}},\ }\href {\doibase 10.1103/PhysRevLett.103.063001}
  {\bibfield  {journal} {\bibinfo  {journal} {Phys. Rev. Lett.}\ }\textbf
  {\bibinfo {volume} {103}},\ \bibinfo {pages} {063001} (\bibinfo {year}
  {2009})}\BibitemShut {NoStop}%
\bibitem [{\citenamefont {Yi}\ \emph {et~al.}(2008)\citenamefont {Yi},
  \citenamefont {Daley}, \citenamefont {Pupillo},\ and\ \citenamefont
  {Zoller}}]{yi08}%
  \BibitemOpen
  \bibfield  {author} {\bibinfo {author} {\bibfnamefont {W.}~\bibnamefont
  {Yi}}, \bibinfo {author} {\bibfnamefont {A.~J.}\ \bibnamefont {Daley}},
  \bibinfo {author} {\bibfnamefont {G.}~\bibnamefont {Pupillo}}, \ and\
  \bibinfo {author} {\bibfnamefont {P.}~\bibnamefont {Zoller}},\ }\href
  {\doibase 10.1088/1367-2630/10/7/073015} {\bibfield  {journal} {\bibinfo
  {journal} {New J. Phys.}\ }\textbf {\bibinfo {volume} {10}},\ \bibinfo
  {pages} {073015} (\bibinfo {year} {2008})}\BibitemShut {NoStop}%
\bibitem [{\citenamefont {Kotochigova}\ \emph {et~al.}(2009)\citenamefont
  {Kotochigova}, \citenamefont {Zelevinsky},\ and\ \citenamefont {Ye}}]{kot09}%
  \BibitemOpen
  \bibfield  {author} {\bibinfo {author} {\bibfnamefont {S.}~\bibnamefont
  {Kotochigova}}, \bibinfo {author} {\bibfnamefont {T.}~\bibnamefont
  {Zelevinsky}}, \ and\ \bibinfo {author} {\bibfnamefont {J.}~\bibnamefont
  {Ye}},\ }\href {\doibase 10.1103/PhysRevA.79.012504} {\bibfield  {journal}
  {\bibinfo  {journal} {Phys. Rev.~A}\ }\textbf {\bibinfo {volume} {79}},\
  \bibinfo {pages} {012504} (\bibinfo {year} {2009})}\BibitemShut {NoStop}%
\bibitem [{\citenamefont {Jones}\ \emph {et~al.}(2006)\citenamefont {Jones},
  \citenamefont {Tiesinga}, \citenamefont {Lett},\ and\ \citenamefont
  {Julienne}}]{jon06}%
  \BibitemOpen
  \bibfield  {author} {\bibinfo {author} {\bibfnamefont {K.~M.}\ \bibnamefont
  {Jones}}, \bibinfo {author} {\bibfnamefont {E.}~\bibnamefont {Tiesinga}},
  \bibinfo {author} {\bibfnamefont {P.~D.}\ \bibnamefont {Lett}}, \ and\
  \bibinfo {author} {\bibfnamefont {P.~S.}\ \bibnamefont {Julienne}},\ }\href
  {\doibase 10.1103/RevModPhys.78.483} {\bibfield  {journal} {\bibinfo
  {journal} {Rev. Mod. Phys.}\ }\textbf {\bibinfo {volume} {78}},\ \bibinfo
  {pages} {483} (\bibinfo {year} {2006})}\BibitemShut {NoStop}%
\bibitem [{\citenamefont {Vogt}\ \emph {et~al.}(2007)\citenamefont {Vogt},
  \citenamefont {Grain}, \citenamefont {Nazarova}, \citenamefont {Sterr},
  \citenamefont {Riehle}, \citenamefont {Lisdat},\ and\ \citenamefont
  {Tiemann}}]{vog07}%
  \BibitemOpen
  \bibfield  {author} {\bibinfo {author} {\bibfnamefont {F.}~\bibnamefont
  {Vogt}}, \bibinfo {author} {\bibfnamefont {C.}~\bibnamefont {Grain}},
  \bibinfo {author} {\bibfnamefont {T.}~\bibnamefont {Nazarova}}, \bibinfo
  {author} {\bibfnamefont {U.}~\bibnamefont {Sterr}}, \bibinfo {author}
  {\bibfnamefont {F.}~\bibnamefont {Riehle}}, \bibinfo {author} {\bibfnamefont
  {C.}~\bibnamefont {Lisdat}}, \ and\ \bibinfo {author} {\bibfnamefont
  {E.}~\bibnamefont {Tiemann}},\ }\href {\doibase 10.1140/epjd/e2007-00175-8}
  {\bibfield  {journal} {\bibinfo  {journal} {Eur. Phys. J. D}\ }\textbf
  {\bibinfo {volume} {44}},\ \bibinfo {pages} {73} (\bibinfo {year} {2007})},\
  \BibitemShut {NoStop}%
\bibitem [{\citenamefont {Reinaudi}\ \emph {et~al.}(2012)\citenamefont
  {Reinaudi}, \citenamefont {Osborn}, \citenamefont {McDonald}, \citenamefont
  {Kotochigova},\ and\ \citenamefont {Zelevinsky}}]{rei12}%
  \BibitemOpen
  \bibfield  {author} {\bibinfo {author} {\bibfnamefont {G.}~\bibnamefont
  {Reinaudi}}, \bibinfo {author} {\bibfnamefont {C.~B.}\ \bibnamefont
  {Osborn}}, \bibinfo {author} {\bibfnamefont {M.}~\bibnamefont {McDonald}},
  \bibinfo {author} {\bibfnamefont {S.}~\bibnamefont {Kotochigova}}, \ and\
  \bibinfo {author} {\bibfnamefont {T.}~\bibnamefont {Zelevinsky}},\ }\href
  {\doibase 10.1103/PhysRevLett.109.115303} {\bibfield  {journal} {\bibinfo
  {journal} {Phys. Rev. Lett.}\ }\textbf {\bibinfo {volume} {109}},\ \bibinfo
  {pages} {115303} (\bibinfo {year} {2012})}\BibitemShut {NoStop}%
\bibitem [{\citenamefont {Kuwamoto}\ \emph {et~al.}(1999)\citenamefont
  {Kuwamoto}, \citenamefont {Honda}, \citenamefont {Takahashi},\ and\
  \citenamefont {Yabuzaki}}]{kuw99}%
  \BibitemOpen
  \bibfield  {author} {\bibinfo {author} {\bibfnamefont {T.}~\bibnamefont
  {Kuwamoto}}, \bibinfo {author} {\bibfnamefont {K.}~\bibnamefont {Honda}},
  \bibinfo {author} {\bibfnamefont {Y.}~\bibnamefont {Takahashi}}, \ and\
  \bibinfo {author} {\bibfnamefont {T.}~\bibnamefont {Yabuzaki}},\ }\href@noop
  {} {\bibfield  {journal} {\bibinfo  {journal} {Phys. Rev.~A}\ }\textbf
  {\bibinfo {volume} {60}},\ \bibinfo {pages} {R745} (\bibinfo {year}
  {1999})}\BibitemShut {NoStop}%
\bibitem [{\citenamefont {Kitagawa}\ \emph {et~al.}(2008)\citenamefont
  {Kitagawa}, \citenamefont {Enomoto}, \citenamefont {Kasa}, \citenamefont
  {Takahashi}, \citenamefont {Ciury{\l}o}, \citenamefont {Naidon},\ and\
  \citenamefont {Julienne}}]{kit08}%
  \BibitemOpen
  \bibfield  {author} {\bibinfo {author} {\bibfnamefont {M.}~\bibnamefont
  {Kitagawa}}, \bibinfo {author} {\bibfnamefont {K.}~\bibnamefont {Enomoto}},
  \bibinfo {author} {\bibfnamefont {K.}~\bibnamefont {Kasa}}, \bibinfo {author}
  {\bibfnamefont {Y.}~\bibnamefont {Takahashi}}, \bibinfo {author}
  {\bibfnamefont {R.}~\bibnamefont {Ciury{\l}o}}, \bibinfo {author}
  {\bibfnamefont {P.}~\bibnamefont {Naidon}}, \ and\ \bibinfo {author}
  {\bibfnamefont {P.~S.}\ \bibnamefont {Julienne}},\ }\href {\doibase
  10.1103/PhysRevA.77.012719} {\bibfield  {journal} {\bibinfo  {journal} {Phys.
  Rev.~A}\ }\textbf {\bibinfo {volume} {77}},\ \bibinfo {pages} {012719}
  (\bibinfo {year} {2008})}\BibitemShut {NoStop}%
\bibitem [{\citenamefont {Kato}\ \emph {et~al.}(2012)\citenamefont {Kato},
  \citenamefont {Yamazaki}, \citenamefont {Shibata}, \citenamefont {Yamamoto},
  \citenamefont {Yamada},\ and\ \citenamefont {Takahashi}}]{kat12}%
  \BibitemOpen
  \bibfield  {author} {\bibinfo {author} {\bibfnamefont {S.}~\bibnamefont
  {Kato}}, \bibinfo {author} {\bibfnamefont {R.}~\bibnamefont {Yamazaki}},
  \bibinfo {author} {\bibfnamefont {K.}~\bibnamefont {Shibata}}, \bibinfo
  {author} {\bibfnamefont {R.}~\bibnamefont {Yamamoto}}, \bibinfo {author}
  {\bibfnamefont {H.}~\bibnamefont {Yamada}}, \ and\ \bibinfo {author}
  {\bibfnamefont {Y.}~\bibnamefont {Takahashi}},\ }\href {\doibase
  10.1103/PhysRevA.86.043411} {\bibfield  {journal} {\bibinfo  {journal} {Phys.
  Rev.~A}\ }\textbf {\bibinfo {volume} {86}},\ \bibinfo {pages} {043411}
  (\bibinfo {year} {2012})}\BibitemShut {NoStop}%
\bibitem [{\citenamefont {Stellmer, S.}\ \emph {et~al.}(2012)\citenamefont {Stellmer},
  \citenamefont {Pasquiou, B.}, \citenamefont {Schreck},F.}]{ste12}%
  \BibitemOpen
  \bibfield  {author} {\bibinfo {author} {\bibfnamefont {S.}~\bibnamefont
  {Stellmer}}, \bibinfo {author} {\bibfnamefont {B.}~\bibnamefont {Pasquiou}}, \bibinfo {author} {\bibfnamefont {R.}~\bibnamefont {Grimm}},
  \bibinfo {author} and {\bibfnamefont {F.}~\bibnamefont {Schreck}}} \href {\doibase
  doi:10.1103/PhysRevLett.109.115302} {\bibfield  {journal} {\bibinfo  {journal}
  {Phys. Rev. Lett.}\ }\textbf {\bibinfo {volume} {109}},\ \bibinfo {pages} {115302}
  (\bibinfo {year} {2012})}\BibitemShut {NoStop}%
\bibitem [{\citenamefont {Zelevinsky}\ \emph {et~al.}(2006)\citenamefont
  {Zelevinsky}, \citenamefont {Boyd}, \citenamefont {Ludlow}, \citenamefont
  {Ido}, \citenamefont {Ye}, \citenamefont {Ciury{\l}o}, \citenamefont
  {Naidon},\ and\ \citenamefont {Julienne}}]{zel06}%
  \BibitemOpen
  \bibfield  {author} {\bibinfo {author} {\bibfnamefont {T.}~\bibnamefont
  {Zelevinsky}}, \bibinfo {author} {\bibfnamefont {M.~M.}\ \bibnamefont
  {Boyd}}, \bibinfo {author} {\bibfnamefont {A.~D.}\ \bibnamefont {Ludlow}},
  \bibinfo {author} {\bibfnamefont {T.}~\bibnamefont {Ido}}, \bibinfo {author}
  {\bibfnamefont {J.}~\bibnamefont {Ye}}, \bibinfo {author} {\bibfnamefont
  {R.}~\bibnamefont {Ciury{\l}o}}, \bibinfo {author} {\bibfnamefont
  {P.}~\bibnamefont {Naidon}}, \ and\ \bibinfo {author} {\bibfnamefont {P.~S.}\
  \bibnamefont {Julienne}},\ }\href {\doibase 10.1103/PhysRevLett.96.203201}
  {\bibfield  {journal} {\bibinfo  {journal} {Phys. Rev. Lett.}\ }\textbf
  {\bibinfo {volume} {96}},\ \bibinfo {pages} {203201} (\bibinfo {year}
  {2006})}\BibitemShut {NoStop}%
\bibitem [{\citenamefont {Martinez~de Escobar}\ \emph
  {et~al.}(2008)\citenamefont {Martinez~de Escobar}, \citenamefont {Mickelson},
  \citenamefont {Pellegrini}, \citenamefont {Nagel}, \citenamefont {Traverso},
  \citenamefont {Yan}, \citenamefont {C\^{o}t\'{e}},\ and\ \citenamefont
  {Killian}}]{mar08e}%
  \BibitemOpen
  \bibfield  {author} {\bibinfo {author} {\bibfnamefont {Y.~N.}\ \bibnamefont
  {Martinez~de Escobar}}, \bibinfo {author} {\bibfnamefont {P.~G.}\
  \bibnamefont {Mickelson}}, \bibinfo {author} {\bibfnamefont {P.}~\bibnamefont
  {Pellegrini}}, \bibinfo {author} {\bibfnamefont {S.~B.}\ \bibnamefont
  {Nagel}}, \bibinfo {author} {\bibfnamefont {A.}~\bibnamefont {Traverso}},
  \bibinfo {author} {\bibfnamefont {M.}~\bibnamefont {Yan}}, \bibinfo {author}
  {\bibfnamefont {R.}~\bibnamefont {C\^{o}t\'{e}}}, \ and\ \bibinfo {author}
  {\bibfnamefont {T.~C.}\ \bibnamefont {Killian}},\ }\href {\doibase
  10.1103/PhysRevA.78.062708} {\bibfield  {journal} {\bibinfo  {journal} {Phys.
  Rev.~A}\ }\textbf {\bibinfo {volume} {78}},\ \bibinfo {pages} {062708}
  (\bibinfo {year} {2008})}\BibitemShut {NoStop}%
\bibitem [{\citenamefont {Degenhardt}\ \emph {et~al.}(2005)\citenamefont
  {Degenhardt}, \citenamefont {Stoehr}, \citenamefont {Lisdat}, \citenamefont
  {Wilpers}, \citenamefont {Schnatz}, \citenamefont {Lipphardt}, \citenamefont
  {Nazarova}, \citenamefont {Pottie}, \citenamefont {Sterr}, \citenamefont
  {Helmcke},\ and\ \citenamefont {Riehle}}]{deg05a}%
  \BibitemOpen
  \bibfield  {author} {\bibinfo {author} {\bibfnamefont {C.}~\bibnamefont
  {Degenhardt}}, \bibinfo {author} {\bibfnamefont {H.}~\bibnamefont {Stoehr}},
  \bibinfo {author} {\bibfnamefont {C.}~\bibnamefont {Lisdat}}, \bibinfo
  {author} {\bibfnamefont {G.}~\bibnamefont {Wilpers}}, \bibinfo {author}
  {\bibfnamefont {H.}~\bibnamefont {Schnatz}}, \bibinfo {author} {\bibfnamefont
  {B.}~\bibnamefont {Lipphardt}}, \bibinfo {author} {\bibfnamefont
  {T.}~\bibnamefont {Nazarova}}, \bibinfo {author} {\bibfnamefont {P.-E.}\
  \bibnamefont {Pottie}}, \bibinfo {author} {\bibfnamefont {U.}~\bibnamefont
  {Sterr}}, \bibinfo {author} {\bibfnamefont {J.}~\bibnamefont {Helmcke}}, \
  and\ \bibinfo {author} {\bibfnamefont {F.}~\bibnamefont {Riehle}},\ }\href
  {\doibase 10.1103/PhysRevA.72.062111} {\bibfield  {journal} {\bibinfo
  {journal} {Phys. Rev.~A}\ }\textbf {\bibinfo {volume} {72}},\ \bibinfo
  {pages} {062111} (\bibinfo {year} {2005})}\BibitemShut {NoStop}%
\bibitem [{\citenamefont {Ciury{\l}o}\ \emph {et~al.}(2004)\citenamefont
  {Ciury{\l}o}, \citenamefont {Tiesinga}, \citenamefont {Kotochigova},\ and\
  \citenamefont {Julienne}}]{ciu04}%
  \BibitemOpen
  \bibfield  {author} {\bibinfo {author} {\bibfnamefont {R.}~\bibnamefont
  {Ciury{\l}o}}, \bibinfo {author} {\bibfnamefont {E.}~\bibnamefont
  {Tiesinga}}, \bibinfo {author} {\bibfnamefont {S.}~\bibnamefont
  {Kotochigova}}, \ and\ \bibinfo {author} {\bibfnamefont {P.~S.}\ \bibnamefont
  {Julienne}},\ }\href {\doibase 10.1103/PhysRevA.70.062710} {\bibfield
  {journal} {\bibinfo  {journal} {Phys. Rev.~A}\ }\textbf {\bibinfo {volume}
  {70}},\ \bibinfo {pages} {062710} (\bibinfo {year} {2004})}\BibitemShut
  {NoStop}%
\bibitem [{\citenamefont {Boisseau}\ \emph {et~al.}(2000)\citenamefont
  {Boisseau}, \citenamefont {Audouard}, \citenamefont {Vigu\'e},\ and\
  \citenamefont {Julienne}}]{boi00}%
  \BibitemOpen
  \bibfield  {author} {\bibinfo {author} {\bibfnamefont {C.}~\bibnamefont
  {Boisseau}}, \bibinfo {author} {\bibfnamefont {E.}~\bibnamefont {Audouard}},
  \bibinfo {author} {\bibfnamefont {J.}~\bibnamefont {Vigu\'e}}, \ and\
  \bibinfo {author} {\bibfnamefont {P.~S.}\ \bibnamefont {Julienne}},\ }\href
  {\doibase 10.1103/PhysRevA.62.052705} {\bibfield  {journal} {\bibinfo
  {journal} {Phys. Rev.~A}\ }\textbf {\bibinfo {volume} {62}},\ \bibinfo
  {pages} {052705} (\bibinfo {year} {2000})}\BibitemShut {NoStop}%
\bibitem [{\citenamefont {Kraft}\ \emph {et~al.}(2009)\citenamefont {Kraft},
  \citenamefont {Vogt}, \citenamefont {Appel}, \citenamefont {Riehle},\ and\
  \citenamefont {Sterr}}]{kra09}%
  \BibitemOpen
  \bibfield  {author} {\bibinfo {author} {\bibfnamefont {S.}~\bibnamefont
  {Kraft}}, \bibinfo {author} {\bibfnamefont {F.}~\bibnamefont {Vogt}},
  \bibinfo {author} {\bibfnamefont {O.}~\bibnamefont {Appel}}, \bibinfo
  {author} {\bibfnamefont {F.}~\bibnamefont {Riehle}}, \ and\ \bibinfo {author}
  {\bibfnamefont {U.}~\bibnamefont {Sterr}},\ }\href {\doibase
  10.1103/PhysRevLett.103.130401} {\bibfield  {journal} {\bibinfo  {journal}
  {Phys. Rev. Lett.}\ }\textbf {\bibinfo {volume} {103}},\ \bibinfo {pages}
  {130401} (\bibinfo {year} {2009})}\BibitemShut {NoStop}%
\bibitem [{\citenamefont {Nazarova}\ \emph {et~al.}(2006)\citenamefont
  {Nazarova}, \citenamefont {Riehle},\ and\ \citenamefont {Sterr}}]{naz06}%
  \BibitemOpen
  \bibfield  {author} {\bibinfo {author} {\bibfnamefont {T.}~\bibnamefont
  {Nazarova}}, \bibinfo {author} {\bibfnamefont {F.}~\bibnamefont {Riehle}}, \
  and\ \bibinfo {author} {\bibfnamefont {U.}~\bibnamefont {Sterr}},\ }\href
  {\doibase 10.1007/s00340-006-2225-y} {\bibfield  {journal} {\bibinfo
  {journal} {Appl. Phys.~B}\ }\textbf {\bibinfo {volume} {83}},\ \bibinfo
  {pages} {531} (\bibinfo {year} {2006})}\BibitemShut {NoStop}%
\bibitem [{sup()}]{supp}%
  \BibitemOpen
  \href@noop {} {}\bibinfo {note} {See Supplemental Material at [URL will be
  inserted by publisher] for details.}\BibitemShut {Stop}%
\bibitem [{\citenamefont {Beverini}\ \emph {et~al.}(1998)\citenamefont
  {Beverini}, \citenamefont {Maccioni},\ and\ \citenamefont {Strumia}}]{bev98}%
  \BibitemOpen
  \bibfield  {author} {\bibinfo {author} {\bibfnamefont {N.}~\bibnamefont
  {Beverini}}, \bibinfo {author} {\bibfnamefont {E.}~\bibnamefont {Maccioni}},
  \ and\ \bibinfo {author} {\bibfnamefont {F.}~\bibnamefont {Strumia}},\ }\href
  {\doibase 10.1364/JOSAB.15.002206} {\bibfield  {journal} {\bibinfo  {journal}
  {J. Opt. Soc. Am.~B}\ }\textbf {\bibinfo {volume} {15}},\ \bibinfo {pages}
  {2206} (\bibinfo {year} {1998})}\BibitemShut {NoStop}%
\bibitem [{\citenamefont {Jones}\ \emph {et~al.}(1999)\citenamefont {Jones},
  \citenamefont {Lett}, \citenamefont {Tiesinga},\ and\ \citenamefont
  {Julienne}}]{jon99}%
  \BibitemOpen
  \bibfield  {author} {\bibinfo {author} {\bibfnamefont {K.~M.}~\bibnamefont
  {Jones}}, \bibinfo {author} {\bibfnamefont {P.~D.}~\bibnamefont {Lett}},\
  \bibinfo {author} {\bibfnamefont {E.}~\bibnamefont {Tiesinga}},\ and\
  \bibinfo {author} {\bibfnamefont {P.~S.}\ \bibnamefont {Julienne}},\ }\href
  {\doibase 10.1103/PhysRevA.61.012501} {\bibfield  {journal} {\bibinfo
  {journal} {Phys. Rev.~A}\ }\textbf {\bibinfo {volume} {61}},\ \bibinfo
  {pages} {012501} (\bibinfo {year} {1999})}\BibitemShut {NoStop}%
\bibitem [{\citenamefont {Theis}\ \emph {et~al.}(2004)\citenamefont {Theis},
  \citenamefont {Thalhammer}, \citenamefont {Winkler}, \citenamefont {Hellwig},
  \citenamefont {Ruff}, \citenamefont {Grimm},\ and\ \citenamefont {{Hecker
  Denschlag}}}]{the04}%
  \BibitemOpen
  \bibfield  {author} {\bibinfo {author} {\bibfnamefont {M.}~\bibnamefont
  {Theis}}, \bibinfo {author} {\bibfnamefont {G.}~\bibnamefont {Thalhammer}},
  \bibinfo {author} {\bibfnamefont {K.}~\bibnamefont {Winkler}}, \bibinfo
  {author} {\bibfnamefont {M.}~\bibnamefont {Hellwig}}, \bibinfo {author}
  {\bibfnamefont {G.}~\bibnamefont {Ruff}}, \bibinfo {author} {\bibfnamefont
  {R.}~\bibnamefont {Grimm}}, \ and\ \bibinfo {author} {\bibfnamefont
  {J.}~\bibnamefont {{Hecker Denschlag}}},\ }\href {\doibase
  10.1103/PhysRevLett.93.123001} {\bibfield  {journal} {\bibinfo  {journal}
  {Phys. Rev. Lett.}\ }\textbf {\bibinfo {volume} {93}},\ \bibinfo {pages}
  {123001} (\bibinfo {year} {2004})}\BibitemShut {NoStop}%
\bibitem [{\citenamefont {Allard}\ \emph {et~al.}(2005)\citenamefont {Allard},
  \citenamefont {Falke}, \citenamefont {Pashov}, \citenamefont {Dulieu},
  \citenamefont {Kn{\"o}ckel},\ and\ \citenamefont {Tiemann}}]{all05}%
  \BibitemOpen
  \bibfield  {author} {\bibinfo {author} {\bibfnamefont {O.}~\bibnamefont
  {Allard}}, \bibinfo {author} {\bibfnamefont {S.}~\bibnamefont {Falke}},
  \bibinfo {author} {\bibfnamefont {A.}~\bibnamefont {Pashov}}, \bibinfo
  {author} {\bibfnamefont {O.}~\bibnamefont {Dulieu}}, \bibinfo {author}
  {\bibfnamefont {H.}~\bibnamefont {Kn{\"o}ckel}}, \ and\ \bibinfo {author}
  {\bibfnamefont {E.}~\bibnamefont {Tiemann}},\ }\href {\doibase
  10.1140/epjd/e2005-00173-x} {\bibfield  {journal} {\bibinfo  {journal} {Eur.
  Phys. J. D}\ }\textbf {\bibinfo {volume} {35}},\ \bibinfo {pages} {483}
  (\bibinfo {year} {2005})}\BibitemShut {NoStop}%
\bibitem [{Note1()}]{Note1}%
  \BibitemOpen
  \bibinfo {note} {There is an opposite sign definition in that paper for the
  dipole-dipole term compared to the conventional usage, which is applied
  through the present paper}\BibitemShut {NoStop}%
\bibitem [{\citenamefont {Townes}\ and\ \citenamefont
  {Schawlow}(1955)}]{tow55}%
  \BibitemOpen
  \bibfield  {author} {\bibinfo {author} {\bibfnamefont {C.~H.}\ \bibnamefont
  {Townes}}\ and\ \bibinfo {author} {\bibfnamefont {A.~L.}\ \bibnamefont
  {Schawlow}},\ }\href@noop {} {\emph {\bibinfo {title} {Microwave
  Spectroscopy}}}\ (\bibinfo  {publisher} {McGraw-Hill},\ \bibinfo {address}
  {New York},\ \bibinfo {year} {1955})\BibitemShut {NoStop}%
\bibitem [{\citenamefont {Mitroy}\ and\ \citenamefont {Zhang}(2008)}]{mit08}%
  \BibitemOpen
  \bibfield  {author} {\bibinfo {author} {\bibfnamefont {J.}~\bibnamefont
  {Mitroy}}\ and\ \bibinfo {author} {\bibfnamefont {J.-Y.}\ \bibnamefont
  {Zhang}},\ }\href {\doibase 10.1063/1.2841470} {\bibfield  {journal}
  {\bibinfo  {journal} {J. Chem. Phys.}\ }\textbf {\bibinfo {volume} {128}},\
  \bibinfo {pages} {134305} (\bibinfo {year} {2008})}\BibitemShut {NoStop}%
\bibitem [{\citenamefont {Tojo}\ \emph {et~al.}(2006)\citenamefont {Tojo},
  \citenamefont {Kitagawa}, \citenamefont {Enomoto}, \citenamefont {Kato},
  \citenamefont {Takasu}, \citenamefont {Kumakura},\ and\ \citenamefont
  {Takahashi}}]{toj06}%
  \BibitemOpen
  \bibfield  {author} {\bibinfo {author} {\bibfnamefont {S.}~\bibnamefont
  {Tojo}}, \bibinfo {author} {\bibfnamefont {M.}~\bibnamefont {Kitagawa}},
  \bibinfo {author} {\bibfnamefont {K.}~\bibnamefont {Enomoto}}, \bibinfo
  {author} {\bibfnamefont {Y.}~\bibnamefont {Kato}}, \bibinfo {author}
  {\bibfnamefont {Y.}~\bibnamefont {Takasu}}, \bibinfo {author} {\bibfnamefont
  {M.}~\bibnamefont {Kumakura}}, \ and\ \bibinfo {author} {\bibfnamefont
  {Y.}~\bibnamefont {Takahashi}},\ }\href {\doibase
  10.1103/PhysRevLett.96.153201} {\bibfield  {journal} {\bibinfo  {journal}
  {Phys. Rev. Lett.}\ }\textbf {\bibinfo {volume} {96}},\ \bibinfo {pages}
  {153201} (\bibinfo {year} {2006})}\BibitemShut {NoStop}%
\bibitem [{\citenamefont {Koch}(2008)}]{koc08}%
  \BibitemOpen
  \bibfield  {author} {\bibinfo {author} {\bibfnamefont {C.~P.}\ \bibnamefont
  {Koch}},\ }\href {\doibase 10.1103/PhysRevA.78.063411} {\bibfield  {journal}
  {\bibinfo  {journal} {Phys. Rev.~A}\ }\textbf {\bibinfo {volume} {78}},\
  \bibinfo {pages} {063411} (\bibinfo {year} {2008})}\BibitemShut {NoStop}%
\bibitem [{\citenamefont {Ciury{\l}o}\ \emph {et~al.}(2005)\citenamefont
  {Ciury{\l}o}, \citenamefont {Tiesinga},\ and\ \citenamefont
  {Julienne}}]{ciu05}%
  \BibitemOpen
  \bibfield  {author} {\bibinfo {author} {\bibfnamefont {R.}~\bibnamefont
  {Ciury{\l}o}}, \bibinfo {author} {\bibfnamefont {E.}~\bibnamefont
  {Tiesinga}}, \ and\ \bibinfo {author} {\bibfnamefont {P.~S.}\ \bibnamefont
  {Julienne}},\ }\href {\doibase 10.1103/PhysRevA.71.030701} {\bibfield
  {journal} {\bibinfo  {journal} {Phys. Rev.~A}\ }\textbf {\bibinfo {volume}
  {71}},\ \bibinfo {pages} {030701} (\bibinfo {year} {2005})}\BibitemShut
  {NoStop}%
\bibitem [{\citenamefont {Fedichev}\ \emph {et~al.}(1996)\citenamefont
  {Fedichev}, \citenamefont {Reynolds},\ and\ \citenamefont
  {Shlyapnikov}}]{fed96}%
  \BibitemOpen
  \bibfield  {author} {\bibinfo {author} {\bibfnamefont {P.~O.}\ \bibnamefont
  {Fedichev}}, \bibinfo {author} {\bibfnamefont {M.~W.}\ \bibnamefont
  {Reynolds}}, \ and\ \bibinfo {author} {\bibfnamefont {G.~V.}\ \bibnamefont
  {Shlyapnikov}},\ }\href {\doibase 10.1103/PhysRevLett.77.2921} {\bibfield
  {journal} {\bibinfo  {journal} {Phys. Rev. Lett.}\ }\textbf {\bibinfo
  {volume} {77}},\ \bibinfo {pages} {2921} (\bibinfo {year}
  {1996})}\BibitemShut {NoStop}%
\bibitem [{\citenamefont {Blatt}\ \emph {et~al.}(2011)\citenamefont {Blatt},
  \citenamefont {Nicholson}, \citenamefont {Bloom}, \citenamefont {Williams},
  \citenamefont {Thomsen}, \citenamefont {Julienne},\ and\ \citenamefont
  {Ye}}]{bla11}%
  \BibitemOpen
  \bibfield  {author} {\bibinfo {author} {\bibfnamefont {S.}~\bibnamefont
  {Blatt}}, \bibinfo {author} {\bibfnamefont {T.~L.}\ \bibnamefont
  {Nicholson}}, \bibinfo {author} {\bibfnamefont {B.~J.}\ \bibnamefont
  {Bloom}}, \bibinfo {author} {\bibfnamefont {J.~R.}\ \bibnamefont {Williams}},
  \bibinfo {author} {\bibfnamefont {J.~W.}\ \bibnamefont {Thomsen}}, \bibinfo
  {author} {\bibfnamefont {P.~S.}\ \bibnamefont {Julienne}}, \ and\ \bibinfo
  {author} {\bibfnamefont {J.}~\bibnamefont {Ye}},\ }\href {\doibase
  10.1103/PhysRevLett.107.073202} {\bibfield  {journal} {\bibinfo  {journal}
  {Phys. Rev. Lett.}\ }\textbf {\bibinfo {volume} {107}},\ \bibinfo {pages}
  {073202} (\bibinfo {year} {2011})}\BibitemShut {NoStop}%
\bibitem [{\citenamefont {Yan}\ \emph {et~al.}(2013)\citenamefont {Yan},
  \citenamefont {DeSalvo}, \citenamefont {Ramachandhran}, \citenamefont {Pu},\
  and\ \citenamefont {Killian}}]{yan13}%
  \BibitemOpen
  \bibfield  {author} {\bibinfo {author} {\bibfnamefont {M.}~\bibnamefont
  {Yan}}, \bibinfo {author} {\bibfnamefont {B.~J.}\ \bibnamefont {DeSalvo}},
  \bibinfo {author} {\bibfnamefont {B.}~\bibnamefont {Ramachandhran}}, \bibinfo
  {author} {\bibfnamefont {H.}~\bibnamefont {Pu}}, \ and\ \bibinfo {author}
  {\bibfnamefont {T.~C.}\ \bibnamefont {Killian}},\ }\href {\doibase
  10.1103/PhysRevLett.110.123201} {\bibfield  {journal} {\bibinfo  {journal}
  {Phys. Rev. Lett.}\ }\textbf {\bibinfo {volume} {110}},\ \bibinfo {pages}
  {123201} (\bibinfo {year} {2013})}\BibitemShut {NoStop}%
\bibitem [{\citenamefont {Enomoto}\ \emph {et~al.}(2008)\citenamefont
  {Enomoto}, \citenamefont {Kasa}, \citenamefont {Kitagawa},\ and\
  \citenamefont {Takahashi}}]{eno08a}%
  \BibitemOpen
  \bibfield  {author} {\bibinfo {author} {\bibfnamefont {K.}~\bibnamefont
  {Enomoto}}, \bibinfo {author} {\bibfnamefont {K.}~\bibnamefont {Kasa}},
  \bibinfo {author} {\bibfnamefont {M.}~\bibnamefont {Kitagawa}}, \ and\
  \bibinfo {author} {\bibfnamefont {Y.}~\bibnamefont {Takahashi}},\ }\href
  {\doibase 10.1103/PhysRevLett.101.203201} {\bibfield  {journal} {\bibinfo
  {journal} {Phys. Rev. Lett.}\ }\textbf {\bibinfo {volume} {101}},\ \bibinfo
  {pages} {203201} (\bibinfo {year} {2008})}\BibitemShut {NoStop}%
\bibitem [{\citenamefont {Rapp}\ \emph {et~al.}(2012)\citenamefont {Rapp},
  \citenamefont {Deng},\ and\ \citenamefont {Santos}}]{rap12}%
  \BibitemOpen
  \bibfield  {author} {\bibinfo {author} {\bibfnamefont {{\'A}.}~\bibnamefont
  {Rapp}}, \bibinfo {author} {\bibfnamefont {X.}~\bibnamefont {Deng}}, \ and\
  \bibinfo {author} {\bibfnamefont {L.}~\bibnamefont {Santos}},\ }\href
  {\doibase 10.1103/PhysRevLett.109.203005} {\bibfield  {journal} {\bibinfo
  {journal} {Phys. Rev. Lett.}\ }\textbf {\bibinfo {volume} {109}},\ \bibinfo
  {pages} {203005} (\bibinfo {year} {2012})}\BibitemShut {NoStop}%
\bibitem [{\citenamefont {Yamazaki}\ \emph {et~al.}(2010)\citenamefont
  {Yamazaki}, \citenamefont {Taie}, \citenamefont {Sugawa},\ and\ \citenamefont
  {Takahashi}}]{yam10}%
  \BibitemOpen
  \bibfield  {author} {\bibinfo {author} {\bibfnamefont {R.}~\bibnamefont
  {Yamazaki}}, \bibinfo {author} {\bibfnamefont {S.}~\bibnamefont {Taie}},
  \bibinfo {author} {\bibfnamefont {S.}~\bibnamefont {Sugawa}}, \ and\ \bibinfo
  {author} {\bibfnamefont {Y.}~\bibnamefont {Takahashi}},\ }\href {\doibase
  10.1103/PhysRevLett.105.050405} {\bibfield  {journal} {\bibinfo  {journal}
  {Phys. Rev. Lett.}\ }\textbf {\bibinfo {volume} {105}},\ \bibinfo {pages}
  {050405} (\bibinfo {year} {2010})}\BibitemShut {NoStop}%
\end{thebibliography}

\begin{thebibliography}{3}
\expandafter\ifx\csname natexlab\endcsname\relax\def\natexlab#1{#1}\fi
\expandafter\ifx\csname bibnamefont\endcsname\relax
  \def\bibnamefont#1{#1}\fi
\expandafter\ifx\csname bibfnamefont\endcsname\relax
  \def\bibfnamefont#1{#1}\fi
\expandafter\ifx\csname citenamefont\endcsname\relax
  \def\citenamefont#1{#1}\fi
\expandafter\ifx\csname url\endcsname\relax
  \def\url#1{\texttt{#1}}\fi
\expandafter\ifx\csname urlprefix\endcsname\relax\def\urlprefix{URL }\fi
\providecommand{\bibinfo}[2]{#2}
\providecommand{\eprint}[2][]{\url{#2}}

\bibitem[{\citenamefont{Kraft et~al.}(2009)\citenamefont{Kraft, Vogt, Appel,
  Riehle, and Sterr}}]{kra09}
\bibinfo{author}{\bibfnamefont{S.}~\bibnamefont{Kraft}},
  \bibinfo{author}{\bibfnamefont{F.}~\bibnamefont{Vogt}},
  \bibinfo{author}{\bibfnamefont{O.}~\bibnamefont{Appel}},
  \bibinfo{author}{\bibfnamefont{F.}~\bibnamefont{Riehle}}, \bibnamefont{and}
  \bibinfo{author}{\bibfnamefont{U.}~\bibnamefont{Sterr}},
  \bibinfo{journal}{Phys. Rev. Lett.} \textbf{\bibinfo{volume}{103}},
  \bibinfo{pages}{130401} (\bibinfo{year}{2009}).

\bibitem[{\citenamefont{Jones et~al.}(1999)\citenamefont{Jones, Lett, Tiesinga,
  and Julienne}}]{jon99}
\bibinfo{author}{\bibfnamefont{K.~M.} \bibnamefont{Jones}},
  \bibinfo{author}{\bibfnamefont{P.~D.} \bibnamefont{Lett}},
  \bibinfo{author}{\bibfnamefont{E.}~\bibnamefont{Tiesinga}}, \bibnamefont{and}
  \bibinfo{author}{\bibfnamefont{P.~S.} \bibnamefont{Julienne}},
  \bibinfo{journal}{Phys. Rev.~A} \textbf{\bibinfo{volume}{61}},
  \bibinfo{pages}{012501} (\bibinfo{year}{1999}).

\bibitem[{\citenamefont{Beverini et~al.}(1998)\citenamefont{Beverini, Maccioni,
  and Strumia}}]{bev98}
\bibinfo{author}{\bibfnamefont{N.}~\bibnamefont{Beverini}},
  \bibinfo{author}{\bibfnamefont{E.}~\bibnamefont{Maccioni}}, \bibnamefont{and}
  \bibinfo{author}{\bibfnamefont{F.}~\bibnamefont{Strumia}},
  \bibinfo{journal}{J. Opt. Soc. Am.~B} \textbf{\bibinfo{volume}{15}},
  \bibinfo{pages}{2206} (\bibinfo{year}{1998}).

\end{thebibliography}

\end{document}